\documentclass[aps,prl,twocolumn,groupedaddress,superscriptaddress,showpacs,longbibliography]{revtex4-1}
\usepackage{amsmath}
\usepackage{graphicx}
\usepackage{float}
\usepackage{xcolor}
\usepackage{textcomp}

\begin{document}

  \title {How to design cell-mediated self-assembled colloidal scaffolds}

  \author{C. S. Dias}
   \email{csdias@fc.ul.pt}
    \affiliation{Departamento de F\'{\i}sica, Faculdade de Ci\^{e}ncias, Universidade de Lisboa, 
    1749-016 Lisboa, Portugal}
    \affiliation{Centro de F\'{i}sica Te\'{o}rica e Computacional, Universidade de Lisboa, 
    1749-016 Lisboa, Portugal}

  \author{C. A. Cust\'odio}
   \email{catarinacustodio@ua.pt}
    \affiliation{Department of Chemistry, CICECO, Campus Universit\'{a}rio de Santiago, University of Aveiro, Aveiro 3810-193, Portugal}    

  \author{G. C. Antunes}
   \email{gcantunes@fc.ul.pt}
    \affiliation{Departamento de F\'{\i}sica, Faculdade de Ci\^{e}ncias, Universidade de Lisboa, 
    1749-016 Lisboa, Portugal}
    \affiliation{Centro de F\'{i}sica Te\'{o}rica e Computacional, Universidade de Lisboa, 
    1749-016 Lisboa, Portugal}

  \author{M. M. Telo da Gama}
   \email{mmgama@fc.ul.pt}
    \affiliation{Departamento de F\'{\i}sica, Faculdade de Ci\^{e}ncias, Universidade de Lisboa, 
    1749-016 Lisboa, Portugal}
    \affiliation{Centro de F\'{i}sica Te\'{o}rica e Computacional, Universidade de Lisboa, 
    1749-016 Lisboa, Portugal}

  \author{J. F. Mano}
   \email{jmano@ua.pt}
    \affiliation{Department of Chemistry, CICECO, Campus Universit\'{a}rio de Santiago, University of Aveiro, Aveiro 3810-193, Portugal}        
    
  \author{N. A. M. Ara\'ujo}
   \email{nmaraujo@fc.ul.pt}
    \affiliation{Departamento de F\'{\i}sica, Faculdade de Ci\^{e}ncias, Universidade de Lisboa, 
    1749-016 Lisboa, Portugal}
    \affiliation{Centro de F\'{i}sica Te\'{o}rica e Computacional, Universidade de Lisboa, 
    1749-016 Lisboa, Portugal}
    
  \begin{abstract}
A critical step in tissue engineering is the design and synthesis of 3D biocompatible matrices (scaffolds) to support and guide the proliferation of cells and tissue growth. Most existing techniques rely on the processing of scaffolds under controlled conditions and then implanting them \textit{in vivo}, with questions related to biocompatibility and the implantation process that are still challenging. As an alternative, it was proposed to assemble the scaffolds \textit{in loco} through the self-organization of colloidal particles mediated by cells.  In this study, we combine experiments, particle-based simulations, and mean-field calculations to show that, in general, the size of the self-assembled scaffold scales with the cell-to-particle ratio. However, we found an optimal value of this ratio, for which the size of the scaffold is maximal when cell-cell adhesion is suppressed.  These results suggest that the size and structure of the self-assembled scaffolds may be designed by tuning the adhesion between cells in the colloidal suspension.
  \end{abstract}

  \maketitle

\section{Introduction}
The ultimate goal of tissue engineering is to produce tissues in a controlled manner to either replace damaged organs or to obtain 3D models to perform fundamental laboratory experiments or to screen drugs~\cite{Marx2015}.
Lessons from natural morphogenesis combined with microengineering strategies may be useful towards functional tissue manufacturing~\cite{Laurent2017}. The spontaneous formation of order occurs in the development of biological structures at different lengths scales, including tissues and organs, mediated by autonomous cellular organization. Engineers have been using the deterministic self-organization processes that take place during embryo development and constantly during tissue renewal to design cell-based structures. In particular, the formation of cell-cell contacts in well-designed microenvironments can be used to produce not only well known spheroids (0D) but also objects with higher dimension, such as fibers (1D)~\cite{Sousa2020}, sheets (2D)~\cite{Matsuda2007} and more complex organoids (3D)~\cite{Takebe2019}. To produce large-scale tissues, biomaterials have been used to define the structure of the tissue engineered \textit{in vitro} and to provide unprecedented control over the cells that interact with them~\cite{Darnell2017}. To that end, a critical step is to design and synthesize scaffolds, as synthetic 3D matrices that support and guide tissue growth~\cite{Murugan2007,Rezwan2006,Hollister2005}. 

\begin{figure}[t]
\begin{center} 
\includegraphics[width=\columnwidth]{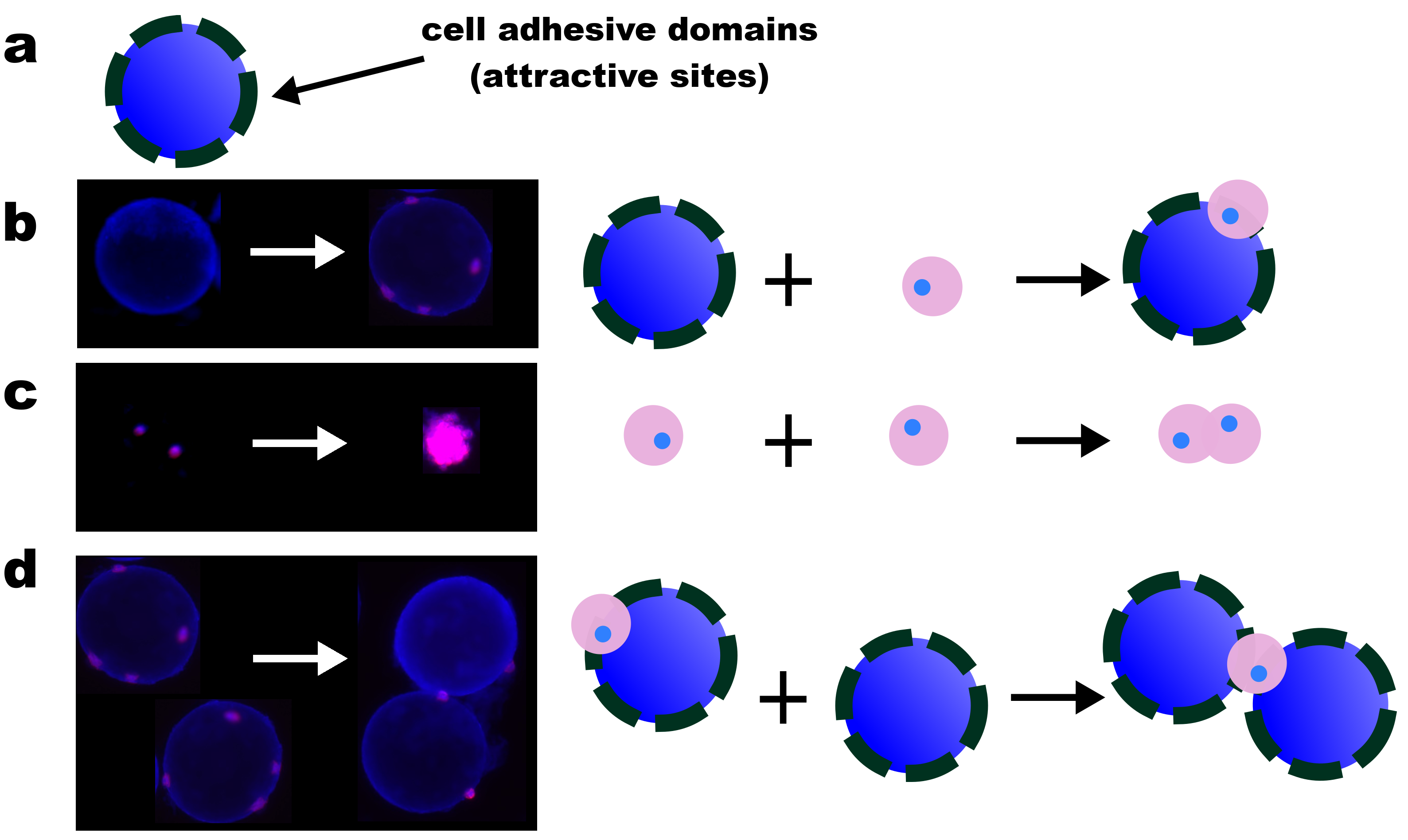} \\
\end{center}
\caption{\textbf{Schematic representation of the relevant processes.} In the experiments, colloidal particles are coated with cell adhesive domains that interact selectively with the cells. \textbf{a.} In the numerical model, the effect of these antibodies is described by a set of (six) attractive (green) sites distributed on the surface of (blue) spherical particles.  
There are three relevant experimental observations (left images taken from different frames of the same experiment): \textbf{b}. cells adhere to the surface of colloidal particles or \textbf{c}. to other cells; \textbf{d}. when a cell adheres to two particles, it mediates a particle-particle bond (stable over the time-scale of the experiment as discussed in the text).~\label{fig.model}}
\end{figure}

From 3D printing to fiber bonding, phase separation, and melt-based technologies, impressive techniques have been developed to assemble the scaffolds, including structures with well-defined architectures~\cite{Khan2015a,Roseti2017}. However, all of them require implanting the final scaffold \textit{in vivo}, which may lead to a number of complications, including expensive, invasive, and risky surgeries~\cite{OBrien2011}. Alternatively, it has been proposed to assemble the scaffolds \textit{in loco} spontaneously from the self-assembly of a suspension of colloidal particles~\cite{Neto2019}. Ideally, this strategy “only” requires injecting a specific colloidal suspension into the damaged area~\cite{Temenoff2000,Oliveira2011}. To control the self-assembly process, Custódio \textit{et al.} devised a protocol to coat the surface of the colloidal particles with bioactive signals. In this way, the attractive interaction between the colloidal particles is mediated by the cells that will promote the spontaneous assembly of an hybrid tissue~\cite{Custodio2014,Custodio2015} (see Fig.~\ref{fig.model}). We hypothesize that such moldable structures could be more effective in generating high-quality tissues because cells will have complete 3D-freedom to mediate the organization of the final construct. The final hybrid tissue will have an amorphous-like structure. However, we can have some control on the size of the agglomerates that are developed and on the level of compaction by changing some parameters, such as geometrical features of the particles, the cell-to-particle ratio and the type of cell-particle interaction. This will be paramount in vital aspects such as diffusion of nutrients or vascularization development. The rationalization of the progress and of the organization of such complex structures cannot be obtained analytically and we propose to explore for the first time the use of computational modeling tools for such an investigation.

In colloidal science, the self-assembly of colloidal structures mediated by a second species (the linkers) is a topic that has attracted sustained interest in recent years~\cite{Peng2016,Joshi2016,Antunes2019, Muller2014,Bharti2014}. A body of theoretical and experimental studies shows that the phase diagrams are enriched by the presence of the linkers~\cite{Lindquist2016,Singh2015,Chen2015,Luo2015}. However, there are two important features that have been neglected in previous studies, which are key to understanding cell-mediated assembly. First, the focus of previous studies has been the structure of the aggregated colloids neglecting at large the effect of the interaction between the linkers \cite{Lowensohn2019,Antunes2019}. In what follows, we consider the cell-cell interaction and show that it plays an important role in determining the final structure. Second, the research on linkers has focused mostly on equilibrium properties based on the assumption that the particle-linker interaction is reversible~\cite{Cyron2013,Lindquist2016}. By contrast, in the cell-mediated assembly, the (cell-particle) adhesion forces are in the range 50 – 500 pm~\cite{Johnson2011}, three to four orders of magnitude larger than the thermal forces. Thus, the cell-particle adhesion is resilient to thermal fluctuations and is practically irreversible in the time scale of the scaffold assembly (~20 hours)~\cite{Custodio2015}. The final colloidal scaffold is then a kinetic structure that depends strongly on the kinetic pathway of the assembly, as well as on the properties of the relevant assembling elements, including the relative number of cells and particles.

\begin{figure*}[t]
   \begin{center} \includegraphics[width=\textwidth]{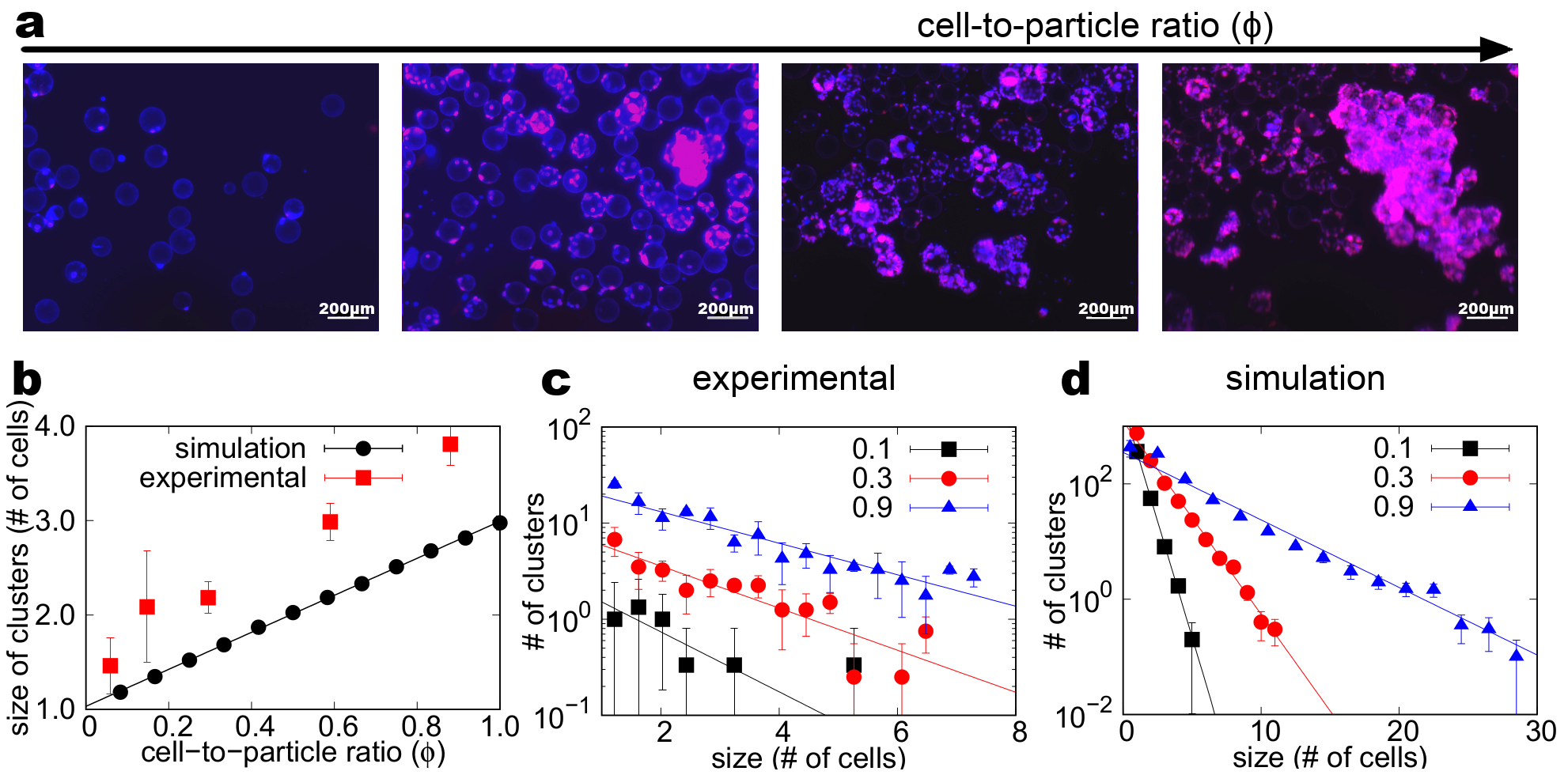} \\
\caption{\textbf{Experimental and particle-based simulation results for the cell-mediated self-assembly of scaffolds.} \textbf{a.} Images obtained experimentally for different cell-to-particle ratios $\phi$, namely, $0.005$, $0.1$, $0.3$, and $0.9$ (defined by the ratio of area covered by all cells if adhered to the surface of particles, as given by Eq.~\eqref{eq.phi}). Particles (of $115 \mu m$ diameter) are in blue and cells are fluorescently marked in pink. The scale bar at the bottom right of each image is 200$\mu$m. b. Average size of independent clusters of cells obtained experimentally and numerically, defined as the average number of contiguous cells (aggregates of cells adhered to each other and/or to particles). Size distribution of the clusters of cells obtained (c.) experimentally and (d.) numerically, for different values of $\phi$, namely, $0.1$, $0.3$, and $0.9$. The simulated sizes are of the order of the experimental ones. In the experiments, clusters of cells are identified by contiguous clusters of pink/blue fluorescent pixels, scaled by the size of one cell estimated from image processing at low cell-to-particle ratio (see the methods section for more details).}\label{fig.cell_attract}

   \end{center}
\end{figure*}
Scaffolds must be mechanically robust, to be resilient to external perturbations, yet sufficiently porous, to guarantee cell migration/proliferation and the access of vital nutrients to the bulk~\cite{OBrien2011,Murugan2007, Badami2006,Chen2007,Li2001,Rezwan2006,Hollister2005}. The spontaneous assembly of cell-mediated colloidal scaffolds entails a very low level of intervention in the process. This is simultaneously an advantage, due to the high scalability, and a challenge, for the level of control over the final structure is limited. The self-assembly process is stochastic in nature and the size and architecture of the scaffold depend on various parameters, such as shape, size, and coating of the colloidal particles, as well as the ratio of the number of cells and particles. A systematic experimental study of these parameters is not feasible as it requires a large number of costly and time consuming experimental cycles \cite{Oliveira2014a}. As an alternative, in what follows, we combine experiments, particle-based simulations, and mean-field calculations, to investigate the role of the cell-to-particle ratio. Under normal conditions, we find that although the number of particle-particle bonds increases with the cell-to-particle ratio, the fraction of cells that mediate bonds does not, as cells adhere not only to the coated particles but also among themselves. However, by blocking cadherins, and suppressing cell-cell adhesion, the bonding efficiency increases for a broad range of the cell-to-particle ratio. This leads to much larger scaffolds for a given cell-to-particle ratio.

\section{Results}
We performed experiments with polystyrene particles decorated with cell adhesive domains, one of the strategies commonly use to increase the bio-instructive character of biomaterials~\cite{Custodio2014a}. In the absence of cells, the particle-particle interaction is mainly repulsive. However, the cells in solution can adhere to the surface of the coated particles. If the same cell adheres to two particles, it mediates a bond between the particles, and promotes cell-mediated particle-particle aggregation (see Fig.~\ref{fig.model}). Thus, the size and structure of the scaffold (which we consider to be the largest aggregate) depends on the number of cell-mediated bonds per particle. To study this dependence in a systematic way, we fixed the number of particles $N_p$ and changed the number of cells $N_c$. For each particle, the maximum number of cell-mediated bonds is given by the number of cells that can cover the surface of the particle, i.e., $A_p/A_c$, where $A_p$ is the average area of the surface of a particle and $A_c$ is the average area of a cell adhered to the surface of a particle (see methods for the explicit values). Thus, in the experiments, we define the cell-to-particle ratio as, 

\begin{equation}\label{eq.phi}
\phi=\frac{N_cA_c}{N_pA_p} \ \ .
\end{equation}

Figure~\ref{fig.cell_attract}(a) contains a set of images obtained experimentally for four different values of $\phi$, where particles are in blue and cells in pink. It is visible that the size of the aggregates of particles, mediated by cells, increases with $\phi$. We identify three relevant processes: cells adhere to the surface of the colloidal particles, adhere to other cells, and mediate particle-particle bonds (see Fig.~\ref{fig.model}).

To model the dynamics numerically, we performed particle-based simulations, where we assumed that each particle can adhere to six cells, i.e., $Ap/Ac = 6$ (experimentally the value may be higher, see methods), and we mimic the effect of the coating by defining six attractive sites on the surface of the spherical colloidal particles. The repulsive particle-particle interaction is described by a Yukawa-like potential and the attractive cell-cell and cell-particle interactions are described by an inverted Gaussian potential (see methods for further details). The stochastic trajectories of the particles are resolved by integrating the corresponding Langevin equation parameterized to yield different diffusion coefficients for the particles $D_p$ and cells $D_c$, namely, $D_c/D_p=10$. In the experiments, particles sediment, and thus we focus on the dynamics close to the substrate. We consider a simulation box of lateral size $L=64$ and height $H=6$, in units of the particle diameter. Particles diffuse
on the substrate (2D) with their equator always parallel to the substrate while cells diffuse in the entire simulation box (3D).

\begin{figure*}[t]
   \begin{center} \includegraphics[width=\textwidth]{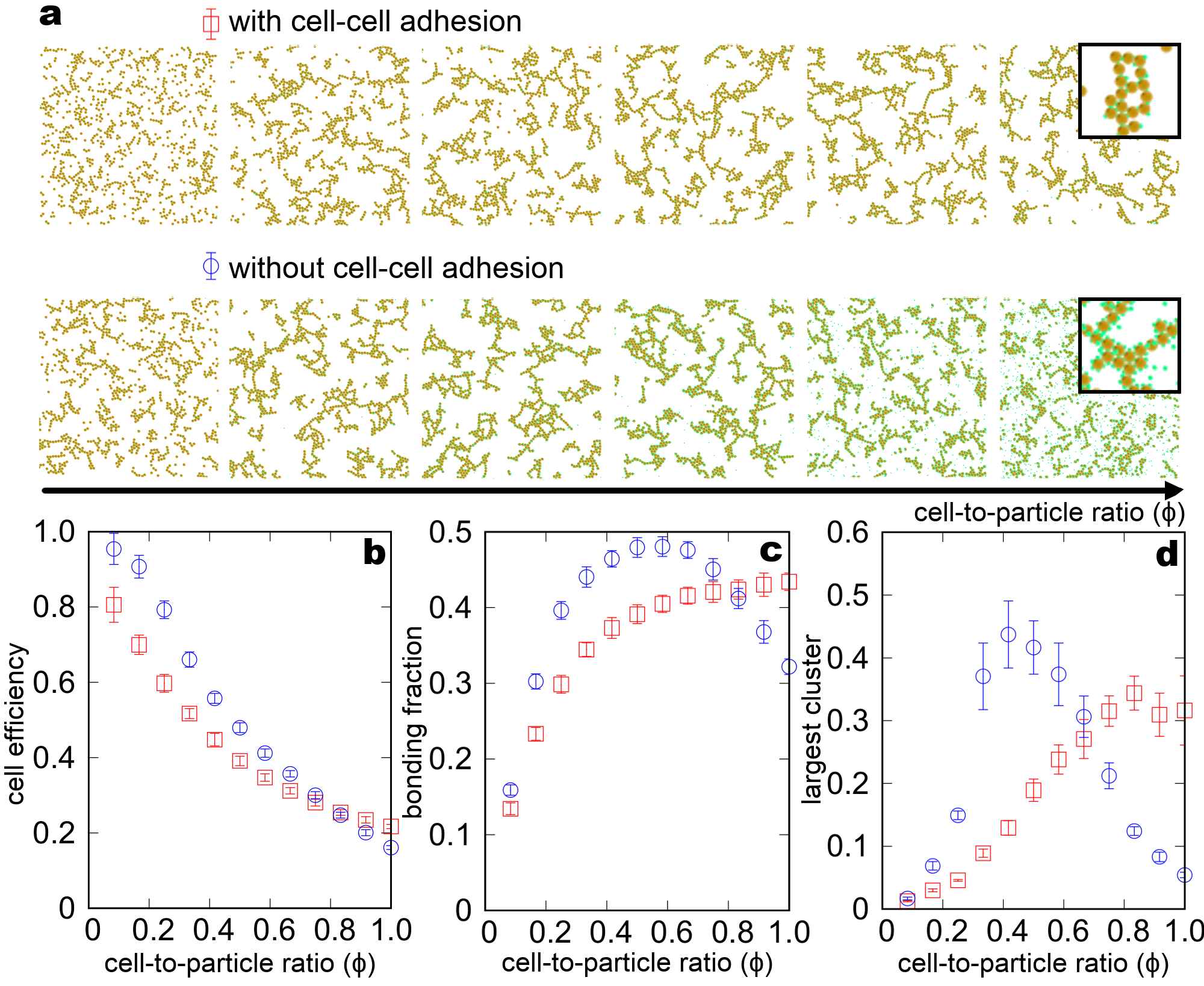} \\
\caption{\textbf{Numerical simulations and mean-field results with and without the cell-cell adhesion mechanism.} \textbf{a.} Snapshots from numerical simulations of particles (yellow) and cells (green) with (top) and without (bottom) cell-cell adhesion for different values of cell-to-particle ratio ($\phi$), namely, $1/6$, $1/3$, $1/2$, $2/3$, $5/6$, and $1$.  Dependence on $\phi$ of \textbf{b.} the cell efficiency, defined as the fraction of particle-particle bonds per cell; \textbf{c.} the bonding fraction, defined as $2N_{b}/6N_p$, where $6N_p$ is the total number of attractive sites ($N_p$ is the number of particles) and $N_{b}$ the number of particle-particle bonds; and \textbf{d.} fraction of particles in the largest aggregate (scaffold). (Red) squares are with cell-cell adhesion and (blue) circles without. Simulations were performed for $1024$ particles in a simulation box of lateral size $L=64$ and height $H=6$ in units of the particle diameter, averaged over $10$ samples.~\label{fig.numerical_phi_effect}}
   \end{center}
\end{figure*}
Cell-cell adhesion occurs naturally both in suspension and on substrates. This can be seen in Fig.~\ref{fig.cell_attract}(a), with several clusters of cells in suspension and on the surface of the particles, for different values of $\phi$. In Fig.~\ref{fig.cell_attract}(b) we show the dependence on $\phi$ of the average size of the clusters of cells obtained numerically and from the experimental images, defined as the number of cells that adhere to each other. The numerical and experimental results are in quantitative agreement and are consistent with a linear increase of the size of the cell clusters with $\phi$. This suggests that, although the size of the scaffold increases with $\phi$, the cell efficiency, defined as the number of particle-particle bonds per cell, decreases. In Fig.~\ref{fig.numerical_phi_effect}(a), we plot this fraction, obtained numerically, revealing a monotonic decrease of the cell efficiency with $\phi$. This decrease is the result of two mechanisms: cell-cell adhesion that promotes the formation of cell clusters and adhesion of cells on the surface of particles in the bulk of the scaffold, where the access of other particles is blocked due to steric effects. To quantify cell-cell adhesion, in Figs.~\ref{fig.cell_attract}(c)~and~(d), we plot the distribution of sizes of the clusters of cells in the experiments and simulations. In both cases, we find an exponential distribution, with a characteristic size that increases with $\phi$, consistent with the increase in the average size observed in Fig.~\ref{fig.cell_attract}(b).
\begin{figure}[ht]
   \begin{center} \includegraphics[width=\columnwidth]{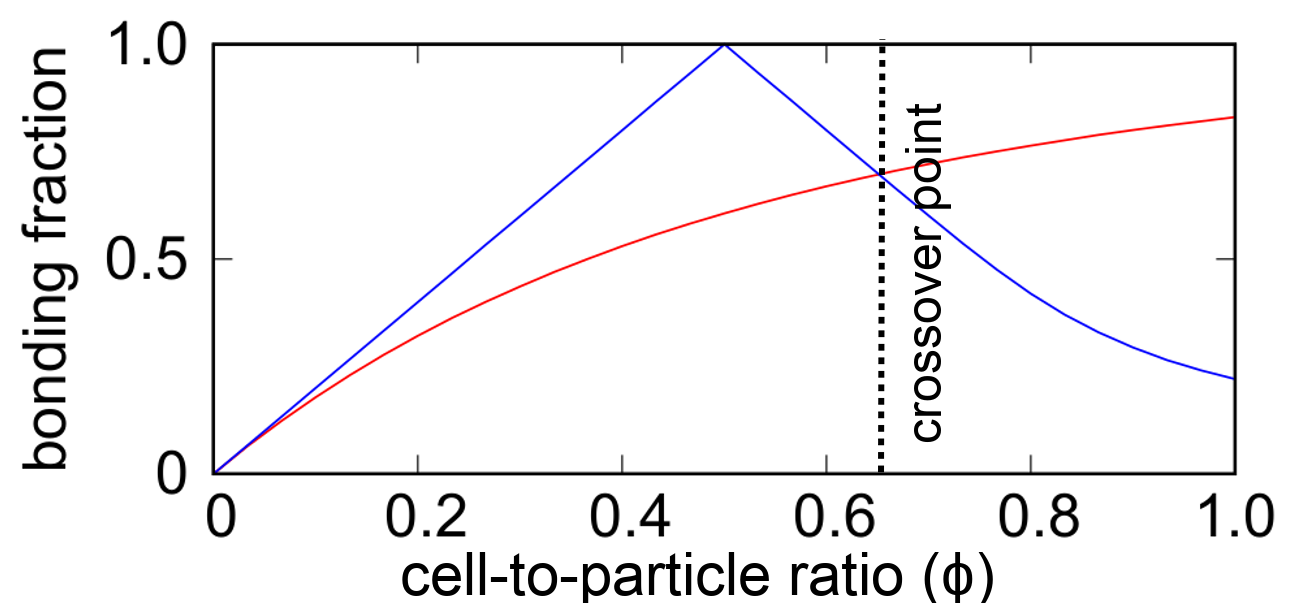} \\
\caption{\textbf{Mean-field results with and without cell-cell adhesion.} Bonding fraction as a function of the cell-to-particle ratio ($\phi$). Results obtained by integrating Eqs.~\eqref{eq.MF}, with $k_0/k_1=10$.  The vertical line is the value of the cell-to-particle ratio above which the size of the scaffold with cell-cell adhesion is larger than the scaffold without cell-cell adhesion.~\label{fig.meanfield_phi_effect}}
   \end{center}
\end{figure}

\textbf{Without cell-cell adhesion.} Since cell-cell adhesion compromises the fraction of particle-particle bonds per cell, we proceed to investigate the dynamics when cell-cell adhesion is suppressed. Experimentally, this is possible by blocking E-cadherins, a transmembrane glycoprotein involved in intercellular adhesion and cytoskeleton-binding functions (see methods for further details). The efficiency of the blocking is shown by a large decrease of cell-cell clusters to a level that they were not observed in the experiments. However, a systematic experimental investigation of this effect is very time and resource demanding. For this reason, we start by investigating it numerically, by performing a new set of simulations where the cell-cell interaction is described by a purely repulsive potential. 

Figure~\ref{fig.numerical_phi_effect}(a) depicts snapshots of the numerical simulations with (top) and without (bottom) cell-cell adhesion, for different values of the cell-to-particle ratio $\phi$. With cell-cell adhesion, the aggregates of particles do increase with $\phi$ for all the values considered. By contrast, without cell-cell adhesion, the characteristic size of the aggregates increases faster with $\phi$ at low $\phi$, and decreases at large $\phi$. This is illustrated in the plot of the size of the largest aggregate of particles (scaffold) as a function of $\phi$, in Fig.~\ref{fig.numerical_phi_effect}(d). With cell-cell adhesion (red squares), the size of the scaffold increases monotonically with $\phi$. By contrast, when cell-cell adhesion is suppressed (blue circles), the size of the scaffold
is maximal at an optimal value of $\phi\approx0.5$. Note that, the size of the scaffold is larger than that of the scaffold with cell-cell adhesion for $\phi<0.6$ and, at the optimal value, the scaffold is twice as large.  This suggests that suppressing cell-cell adhesion is an effective way of optimizing the size of the scaffolds.

Figure~\ref{fig.numerical_phi_effect}(a) illustrates the cell efficiency defined as the number of particle-particle bonds per cell. As expected, this fraction is higher when cell-cell adhesion is suppressed, as there are no clusters of cells in the suspension, but it still decreases monotonically with $\phi$. The increase in cell efficiency justifies the assembly of larger scaffolds without cell-cell adhesion but not the observed non-monotonic behavior. To understand the latter, we measured the bonding fraction, defined as the fraction of sites that are bonded, given by $2N_b/6N_p$, where $N_p$ is the number of particles, $6N_p$ the total number of sites, and $N_b$ the number of cell-mediated bonds.  Figure~\ref{fig.numerical_phi_effect}(c) shows the dependence of the bonding fraction on $\phi$. With cell-cell adhesion, the bonding fraction increases with $\phi$, as a larger number of cells per particle increases the probability of forming a cell-mediated bond. By contrast, in line with the results for the size of the scaffold, without
cell-cell adhesion the bonding fraction is maximal at $\phi=0.5$. At this ratio, the number of cells is half the total number of available sites. 

The position of the maximum for $\phi=0.5$ is justified as follows.  As cells diffuse ten times faster than particles, they will adhere to all the available sites before a significant number of particle-particle bonds is formed. For a cell-mediated bond to form, one needs an occupied site with a cell and an empty one. Assuming a uniform distribution of cells over the available sites, the number of pairs of occupied/empty sites increases with $\phi$ for $\phi<0.5$ and decreases above this optimal value, being a maximum at $\phi=0.5$. Note that the maximum in the bonding fraction at $\phi=0.5$ is also one-half. This means that one-half of the cells mediate bonds while the other half adheres to one particle only. Since the particle-particle bonds are irreversible, asymptotically, there are no free cells, but a large fraction of sites is not available for bonding due to steric effects, even if cells adhere to them. Thus, the bonding fraction is not one but rather one-half. 
The competition between the number of cells available to mediate bonds and the number of pairs of occupied/empty sites is a robust mechanism that should be independent of the spatial dimension. In order to proceed, we propose a set of mean-field rate equations for the density of cells and empty sites. We consider four populations: empty sites, free cells (or clusters of cells), cells adhered to a site, and cells adhered to two sites, mediating a bond, with densities $\rho(t)$, $C_0(t)$, $C_1(t)$, and $C_2(t)$, respectively. The time dependence of $C_0(t)$ is given by, 

\begin{equation}\label{eq:c0}
\dot{C_0}=-k_0 C_0 \rho - \alpha k_0 C_0 C_1 - \alpha k_2 C_0^2,\\
\end{equation} 
where, on the right-hand side, the terms correspond to the adhesion of a free cell to an empty site (first term), to a cell adhered to a site (second term), or to another free cell (third term). $k_0$ and $k_1$ are reaction rates that depend on the diffusion coefficients of cells and particles, which we assume constant. $\alpha$ is a parameter equal to one with cell-cell adhesion and zero when this mechanism is suppressed.  In the same way, the time dependence of the other densities is given by,

\begin{eqnarray}
\dot{C_1}&=&k_0 C_0 \rho - k_1 C_1 \rho -\alpha k_1 C_1^2,\nonumber\\
\dot{C_2}&=&k_1 C_1 \rho + \frac{1}{2}\alpha k_1 C_1^2,\label{eq.MF}\\
\dot{\rho}&=&-k_0 C_0 \rho -k_1 C_1 \rho. \nonumber
\end{eqnarray}

Figure~\ref{fig.meanfield_phi_effect} shows the dependence on $C_0(0)/\rho(0)$ (equal to $\phi$ the cell-to-particle ratio) of the asymptotic bonding fraction ($2C_2(t)/\rho(0)$), obtained for $k_0/k_1=10$, with ($\alpha=0$) and without ($\alpha=1$) cell-cell adhesion. The results clearly reveal the monotonic increase of the bonding fraction when cell-cell adhesion is considered, and the maximum at $\phi=0.5$ when this mechanism is suppressed. However, the bonding fraction at the maximum is one, since in mean-field there are no steric effects.  It is also noteworthy that the two lines cross at a value of $\phi\in[0.6,0.7]$ as observed  in the numerical simulations.

\begin{figure}[t]
   \begin{center} \includegraphics[width=\columnwidth]{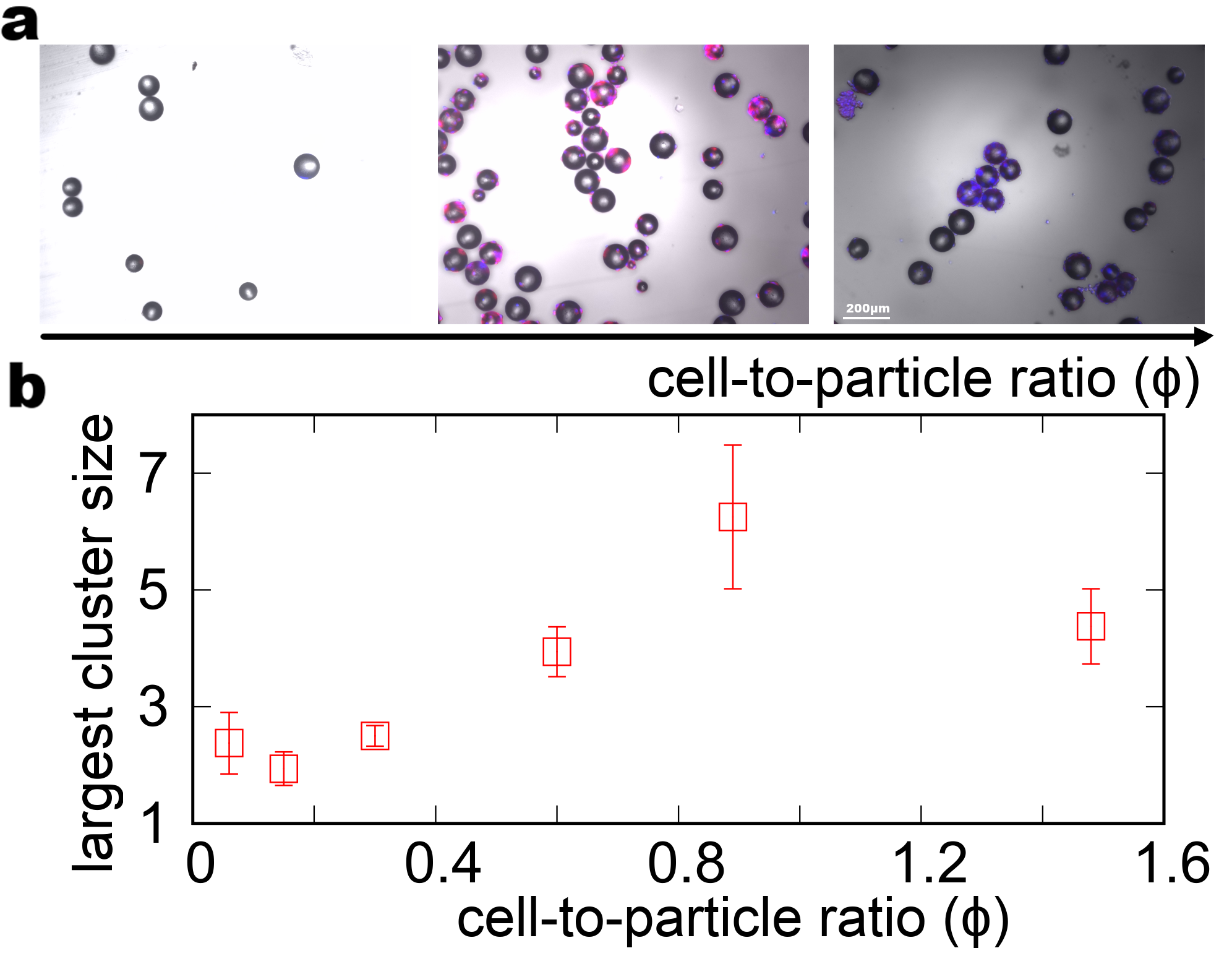} \\
\caption{\textbf{Experimental results without cell-cell adhesion.} \textbf{a.} Experimental images without cell-cell adhesion as a function of the cell-to-particle ratio, namely, $0.06$, $0.9$, and $1.5$. \textbf{b.} Average number of particles in the largest cluster (scaffold) as a function of the cell-to-particle ratio. Results are averages over equal-sized square regions of one experimental image obtained by dividing the image into $16$ squares.~\label{fig.experimental_phi_effect}}
   \end{center}
\end{figure}

The numerical and analytical results suggest that, in the absence of cell-cell adhesion, the size of the scaffold is a non-monotonic function of the cell-to-particle ratio. To verify this prediction, we performed experiments where E-cadherins are blocked (see methods for further details) for different values of $\phi$. In Fig.~\ref{fig.experimental_phi_effect}(a) are the experimental images, for three values of $\phi$. It is clear from the images that the size of the aggregate of particles is larger for the intermediate value of $\phi$.  Figure~\ref{fig.experimental_phi_effect}(b) shows the number of particles in the largest aggregate (scaffold) as a function of $\phi$. As predicted, a maximum is observed in the experiments. The results also reveal a shift of the optimal value of $\phi$ to larger values. This shift may be due to the way $\phi$ is computed in the experiments. We assumed that about twelve cells can adhere to the surface of each particle, based on the relation between the area of a cell and that of a particle. However, even if this is the case, the number of possible bonds per particle is limited by spatial constraints (on a substrate, this is of the order of six as in the simulations), what compromises the cell efficiency and might be responsible for the shift in the optimal value of $\phi$. Nevertheless, the non-monotonic
behavior is confirmed, which is in sharp contrast to what is observed in systems with cell-cell adhesion.

\section{Discussion} 
We compare the efficiency of the cell-mediated self-assembly of colloidal scaffolds with and without cell-cell adhesion. With cell-cell adhesion, the size of the scaffolds increases with the cell-to-particle ratio ($\phi$) as expected, as more cells are available to mediate particle-particle bonds. However, the fraction of bonds per cell decreases with $\phi$, since clusters of cells are formed in the suspension and the number of bonds per particle is reduced by steric effects. When cell-cell adhesion is suppressed, the fraction of bonds per cell at the same value of $\phi$ increases, leading to larger scaffolds for a wide range of $\phi$. By contrast, in the absence of cell-cell adhesion, the size of the largest aggregate (scaffold) is a non-monotonic function of $\phi$, with a maximum at $\phi\approx0.5$. We have shown that this maximum stems from the competition between the number of cells available to mediate bonds and the number of available pairs of occupied/empty sites. This is a very robust mechanism captured by a mean-field approximation and observed experimentally.

The advantages of assembling scaffolds spontaneously \textit{in loco} are many, in particular when it comes to biocompatibility and implantation or time costs. However, the level of control over the final structure is significantly compromised by the stochastic nature of the self-assembly process. Our work reveals that the size of the scaffolds may be increased and controlled by blocking E-cadherins during the initial stage of the aggregation. This blocking strategy does not affect the genetic material of the cell and thus, the offspring will not be blocked, as required for the growth of healthy tissue. To obtain these findings we combined experiments, simulations, and mean-field calculations. Through this combined strategy, it was possible to develop and parameterize a simple theoretical model, in order to explore the parameter space in an effective way. We succeeded in identifying ways to optimize the design of moldable hybrid self-assembled scaffolds for tissue engineering, which were confirmed by experiment. The tools explored in this work may be very useful for describing and predicting the overall spatiotemporal structural development of self-assembled hybrid constructs. A better control of such synthetic morphogenesis mediated by cell-cell and cell-biomaterial interactions could lead to more viable and better-defined tissues to be applied in therapies and as models for drug screening or for fundamental biological studies.

\section{Methods}

\textbf{Experimental cell culture} Experimental cell culture tests were performed using L929 mouse fibroblast cell line (European Collection of Authenticated Cell Cultures). L929 is a very well characterized cell line and the reference used in cytotoxicity and screening assays. L929 cells were cultured in Dulbeccos modified Eagle medium low glucose (Sigma-Aldrich), supplemented with $10\%$ fetal bovine serum (FBS, Thermo Fisher Scientic) and $1\%$ antibiotic/antimycotic (Thermo Fisher Scientic).

To block E-cadherins, the cells were incubated for 30 min with CD324 antibody, using $2.0 \mu g$ of antibody per $5 \times 105$ cells in $100 \mu l$ volume. This is a fourfold increase in concentration of antibody per million cells, by comparison to the standard protocol for cell labelling. After blocking, the cells were centrifuged and washed with PBS before seeding.

\textbf{Experimental apparatus.} Polystyrene microparticles (Polysciences, Inc) were used as support for cell attachment. Microparticles were pretreated with oxygen plasma, for oxidizing the surface and increasing the hydrophilicity followed by a coating with an adhesive protein. Briefly, the particles were treated with ATTO low-pressure plasma system (Diener) for 10 min (30 V, 0.6 mbar), sterilized with ethanol 70$\%$ and dried. The sterilized particles where incubated in a solution of fibronectin (20ug/ml) for 1hour and washed with PBS.  To block E-cadherins, cells were incubated for 30min with CD324 antibody, using 2.0 $\mu g$ per 5x105 cells in 100 $\mu l$ volume, centrifuged and washed with PBS.  Normal cells and cells with blocked cadherins were seeded in the microparticles in a 8-well plate (Ibidi) and cultured up to 24 hours at 37Â°C in a humidified 5$\%$ $CO_2$ air atmosphere. The cytoskeleton of the cells cultured in the microparticles was visualized after fluorescence phalloidin staining. Samples were incubated in Flash
Phalloidin Red 594 (1:40 in PBS, Biolegend) for 45 min at 37ÂºC. DNA was stained with DAPI (1:1000, 1 $mg.mL^{-1}$ in PBS, ThermoFisher Scientific) for 5 min. Samples were visualized by fluorescence microscopy (Axio Imager 2, Zeiss).

\textbf{Image processing.} On Fig.~\ref{fig.cell_attract}(a), we can see color microscopy images of the experimental scaffold formation where cells are marked with phalloidin/DAPI. To perform image recognition of the cells we use the python package scikit-image \cite{VanderWalt2014}. We start by converting the color space from RGB to the CIE 1931 color space in order to separate the blue color (typical of the particle) from the fluorescent pink (typical of the cells) and then convert to an intensity scale (gray scale). We then filter the intensity scale image to recognize only the cells clusters by applying a combination of a local (weighted mean from a local neighborhood of a pixel) and a global (Otsu's method \cite{Otsu1979}) threshold to the pixels. Properties of the clusters of pixels are then measured.  To measure the size of cell clusters, we performed image recognition on the experimental microscopic images (see details in the methods section). Since cells can have diameters from around 10 microns in suspension to 50 microns for spread cells, we measured directly the size of a cell at the lowest cell-to-particle ratio
experimentally measured. Using such low density of cells allows to attempt at a calibration of the size of a single cell. We have measured that single cells can vary in size from 150$\pm$10 to 300$\pm$10 pixels. Where above that limit pixel clusters are cell clusters.

\textbf{Numerical simulations} Simulations of Langevin dynamics were performed for particles with six equally spaced sites along their equator, which interact attractively with cells (we performed also simulations with six sites
equally spaced over the entire particle surface and the results were consistent with those reported here). As in Refs.~\cite{Dias2016,Araujo2017}, the translational and rotational motion of the particles are described by the following Langevin equations, 
\begin{equation}
 m\dot{\vec{v_i}}(t)=-\nabla_{\vec{r_i}}
U-\frac{m}{\tau_t}\vec{v_i}(t)+\sqrt{\frac{2mk_BT}{\tau_t}}\vec{\xi_t^i}(t),
\label{eq.Langevin_dynamics_trans} 
\end{equation}
and
\begin{equation}
 I\dot{\vec{\omega_i}}(t)=-\nabla_{\vec{\theta_i}}
U-\frac{I}{\tau_r}\vec{\omega_i}(t)+\sqrt{\frac{2Ik_BT}{\tau_r}}\vec{\xi_r^i}(t),
\label{eq.Langevin_dynamics_rot}
\end{equation}
where, $\vec{v_i}$ and $\vec{\omega_i}$ are the translational and angular
velocities of particle $i$. Particles are spherical with mass $m$ and inertia
$I$ and the sites on their surface have negligible mass. $\tau_t$ and $\tau_r$
are the translational and rotational damping times (for spherical particles
$\tau_r=10\tau_t/3$). $\vec{\xi_t^i}(t)$ and $\vec{\xi_r^i}(t)$ are stochastic
terms, and $U$ is the total potential with contributions from the
particle-particle interactions. These equations were integrated using the
velocity Verlet scheme, implemented in the Large-scale Atomic/Molecular
Massively Parallel Simulator (LAMMPS)~\cite{Plimpton1995}. 

We consider a repulsive Yukawa-like potential for the particle-particle
interaction, given by,
\begin{equation}
U_{\text{part/part}}(r)=\frac{A}{k}\exp{\left[-k\left(r-d_p\right)\right]},
\label{eq.yukawa}
\end{equation}
where $A/k=k_BT/4$ is the energy scale and $k$ the screening length. We consider a
Gaussian attractive potential for the site-cell interaction~\cite{Dias2016},
given by,
\begin{equation}
U_{\text{site/cell}}(r_p)=-\epsilon\exp\left[-(r_p/\sigma)^2\right],
\label{eq.gaussian} 
\end{equation}
where $r_p$ is the distance between the center of the site and the cell,
$\epsilon=20k_BT$ is the interaction strength that sets the energy scale, and
$\sigma$ the width of the Gaussian. We consider two types of cell-cell
interaction, attractive (to mimic cell-cell adhesion) and repulsive (to mimic
excluded volume, when we block the cell-cell adhesion).

Since particles sediment, we consider a flat (attractive) substrate with an
initial coverage similar to the experimental one (the sizes of the simulated
systems are also of the order of the experimental ones). Note that, although
the particles are on a planar substrate, they can still rotate in three
dimensions. 

\textbf{Cell-to-particle ratio.} To compute the cell-to-particle ratio,
$\phi$, defined in Eq.~\eqref{eq.phi}, we considered $N_p= 900$ in
Fig.~\ref{fig.cell_attract} and $N_p= 800$ in
Fig.~\ref{fig.experimental_phi_effect}. For the surface area of the particles,
we considered $A_p=200^2\pi\mu m^2$ in Fig.~\ref{fig.cell_attract} and $A_p=
115^2\pi \mu m^2$ in Fig.~\ref{fig.experimental_phi_effect}. The typical diameter of a cell fully spread on the surface of a particle is $50\mu m$, and we assumed a circular shape and used $A_c=25^2\pi \mu m^2$. This contrasts to the diameter of a cell in suspension of around $10 \mu m$, which we considered in the relation between the cell and particle diffusion coefficients in the numerical simulations.

\bibliography{library}

\begin{thebibliography}{39}%
\makeatletter
\providecommand \@ifxundefined [1]{%
 \@ifx{#1\undefined}
}%
\providecommand \@ifnum [1]{%
 \ifnum #1\expandafter \@firstoftwo
 \else \expandafter \@secondoftwo
 \fi
}%
\providecommand \@ifx [1]{%
 \ifx #1\expandafter \@firstoftwo
 \else \expandafter \@secondoftwo
 \fi
}%
\providecommand \natexlab [1]{#1}%
\providecommand \enquote  [1]{``#1''}%
\providecommand \bibnamefont  [1]{#1}%
\providecommand \bibfnamefont [1]{#1}%
\providecommand \citenamefont [1]{#1}%
\providecommand \href@noop [0]{\@secondoftwo}%
\providecommand \href [0]{\begingroup \@sanitize@url \@href}%
\providecommand \@href[1]{\@@startlink{#1}\@@href}%
\providecommand \@@href[1]{\endgroup#1\@@endlink}%
\providecommand \@sanitize@url [0]{\catcode `\\12\catcode `\$12\catcode
  `\&12\catcode `\#12\catcode `\^12\catcode `\_12\catcode `\%12\relax}%
\providecommand \@@startlink[1]{}%
\providecommand \@@endlink[0]{}%
\providecommand \url  [0]{\begingroup\@sanitize@url \@url }%
\providecommand \@url [1]{\endgroup\@href {#1}{\urlprefix }}%
\providecommand \urlprefix  [0]{URL }%
\providecommand \Eprint [0]{\href }%
\providecommand \doibase [0]{http://dx.doi.org/}%
\providecommand \selectlanguage [0]{\@gobble}%
\providecommand \bibinfo  [0]{\@secondoftwo}%
\providecommand \bibfield  [0]{\@secondoftwo}%
\providecommand \translation [1]{[#1]}%
\providecommand \BibitemOpen [0]{}%
\providecommand \bibitemStop [0]{}%
\providecommand \bibitemNoStop [0]{.\EOS\space}%
\providecommand \EOS [0]{\spacefactor3000\relax}%
\providecommand \BibitemShut  [1]{\csname bibitem#1\endcsname}%
\let\auto@bib@innerbib\@empty
%</preamble>
\bibitem [{\citenamefont {Marx}(2015)}]{Marx2015}%
  \BibitemOpen
  \bibfield  {author} {\bibinfo {author} {\bibfnamefont {V.}~\bibnamefont
  {Marx}},\ }\bibfield  {title} {\enquote {\bibinfo {title} {{Tissue
  engineering: Organs from the lab}},}\ }\href@noop {} {\bibfield  {journal}
  {\bibinfo  {journal} {Nature}\ }\textbf {\bibinfo {volume} {522}},\ \bibinfo
  {pages} {373} (\bibinfo {year} {2015})}\BibitemShut {NoStop}%
\bibitem [{\citenamefont {Laurent}\ \emph {et~al.}(2017)\citenamefont
  {Laurent}, \citenamefont {Blin}, \citenamefont {Chatelain}, \citenamefont
  {Vanneaux}, \citenamefont {Fuchs}, \citenamefont {Larghero},\ and\
  \citenamefont {Th{\'{e}}ry}}]{Laurent2017}%
  \BibitemOpen
  \bibfield  {author} {\bibinfo {author} {\bibfnamefont {J.}~\bibnamefont
  {Laurent}}, \bibinfo {author} {\bibfnamefont {G.}~\bibnamefont {Blin}},
  \bibinfo {author} {\bibfnamefont {F.}~\bibnamefont {Chatelain}}, \bibinfo
  {author} {\bibfnamefont {V.}~\bibnamefont {Vanneaux}}, \bibinfo {author}
  {\bibfnamefont {A.}~\bibnamefont {Fuchs}}, \bibinfo {author} {\bibfnamefont
  {J.}~\bibnamefont {Larghero}}, \ and\ \bibinfo {author} {\bibfnamefont
  {M.}~\bibnamefont {Th{\'{e}}ry}},\ }\bibfield  {title} {\enquote {\bibinfo
  {title} {{Convergence of microengineering and cellular self-organization
  towards functional tissue manufacturing}},}\ }\href@noop {} {\bibfield
  {journal} {\bibinfo  {journal} {Nat. Biomed. Eng.}\ }\textbf {\bibinfo
  {volume} {1}},\ \bibinfo {pages} {939} (\bibinfo {year} {2017})}\BibitemShut
  {NoStop}%
\bibitem [{\citenamefont {Sousa}\ \emph {et~al.}(2020)\citenamefont {Sousa},
  \citenamefont {Martins-Cruz}, \citenamefont {Oliveira},\ and\ \citenamefont
  {Mano}}]{Sousa2020}%
  \BibitemOpen
  \bibfield  {author} {\bibinfo {author} {\bibfnamefont {A.~R.}\ \bibnamefont
  {Sousa}}, \bibinfo {author} {\bibfnamefont {C.}~\bibnamefont {Martins-Cruz}},
  \bibinfo {author} {\bibfnamefont {M.~B.}\ \bibnamefont {Oliveira}}, \ and\
  \bibinfo {author} {\bibfnamefont {J.~F.}\ \bibnamefont {Mano}},\ }\bibfield
  {title} {\enquote {\bibinfo {title} {{One-Step Rapid Fabrication of Cell-Only
  Living Fibers}},}\ }\href {\doibase 10.1002/adma.201906305} {\bibfield
  {journal} {\bibinfo  {journal} {Adv. Mater.}\ }\textbf {\bibinfo {volume}
  {32}},\ \bibinfo {pages} {1906305} (\bibinfo {year} {2020})}\BibitemShut
  {NoStop}%
\bibitem [{\citenamefont {Matsuda}\ \emph {et~al.}(2007)\citenamefont
  {Matsuda}, \citenamefont {Shimizu}, \citenamefont {Yamato},\ and\
  \citenamefont {Okano}}]{Matsuda2007}%
  \BibitemOpen
  \bibfield  {author} {\bibinfo {author} {\bibfnamefont {N.}~\bibnamefont
  {Matsuda}}, \bibinfo {author} {\bibfnamefont {T.}~\bibnamefont {Shimizu}},
  \bibinfo {author} {\bibfnamefont {M.}~\bibnamefont {Yamato}}, \ and\ \bibinfo
  {author} {\bibfnamefont {T.}~\bibnamefont {Okano}},\ }\bibfield  {title}
  {\enquote {\bibinfo {title} {{Tissue engineering based on cell sheet
  technology}},}\ }\href@noop {} {\bibfield  {journal} {\bibinfo  {journal}
  {Adv. Mater.}\ }\textbf {\bibinfo {volume} {19}},\ \bibinfo {pages} {3089}
  (\bibinfo {year} {2007})}\BibitemShut {NoStop}%
\bibitem [{\citenamefont {Takebe}\ and\ \citenamefont
  {Wells}(2019)}]{Takebe2019}%
  \BibitemOpen
  \bibfield  {author} {\bibinfo {author} {\bibfnamefont {T.}~\bibnamefont
  {Takebe}}\ and\ \bibinfo {author} {\bibfnamefont {J.~M.}\ \bibnamefont
  {Wells}},\ }\bibfield  {title} {\enquote {\bibinfo {title} {{Organoids by
  design}},}\ }\href@noop {} {\bibfield  {journal} {\bibinfo  {journal}
  {Science}\ }\textbf {\bibinfo {volume} {364}},\ \bibinfo {pages} {956}
  (\bibinfo {year} {2019})}\BibitemShut {NoStop}%
\bibitem [{\citenamefont {Darnell}\ and\ \citenamefont
  {Mooney}(2017)}]{Darnell2017}%
  \BibitemOpen
  \bibfield  {author} {\bibinfo {author} {\bibfnamefont {M.}~\bibnamefont
  {Darnell}}\ and\ \bibinfo {author} {\bibfnamefont {D.~J.}\ \bibnamefont
  {Mooney}},\ }\bibfield  {title} {\enquote {\bibinfo {title} {{Leveraging
  advances in biology to design biomaterials}},}\ }\href@noop {} {\bibfield
  {journal} {\bibinfo  {journal} {Nat. Mater.}\ }\textbf {\bibinfo {volume}
  {16}},\ \bibinfo {pages} {1178} (\bibinfo {year} {2017})}\BibitemShut
  {NoStop}%
\bibitem [{\citenamefont {Murugan}\ and\ \citenamefont
  {Ramakrishna}(2007)}]{Murugan2007}%
  \BibitemOpen
  \bibfield  {author} {\bibinfo {author} {\bibfnamefont {R.}~\bibnamefont
  {Murugan}}\ and\ \bibinfo {author} {\bibfnamefont {S.}~\bibnamefont
  {Ramakrishna}},\ }\bibfield  {title} {\enquote {\bibinfo {title} {{Review
  Design Strategies of Tissue Engineering Scaffolds with Controlled Fiber
  Orientation}},}\ }\href@noop {} {\bibfield  {journal} {\bibinfo  {journal}
  {Tissue Eng.}\ }\textbf {\bibinfo {volume} {13}},\ \bibinfo {pages} {1845}
  (\bibinfo {year} {2007})}\BibitemShut {NoStop}%
\bibitem [{\citenamefont {Rezwan}\ \emph {et~al.}(2006)\citenamefont {Rezwan},
  \citenamefont {Chen}, \citenamefont {Blaker},\ and\ \citenamefont
  {Boccaccini}}]{Rezwan2006}%
  \BibitemOpen
  \bibfield  {author} {\bibinfo {author} {\bibfnamefont {K.}~\bibnamefont
  {Rezwan}}, \bibinfo {author} {\bibfnamefont {Q.~Z.}\ \bibnamefont {Chen}},
  \bibinfo {author} {\bibfnamefont {J.~J.}\ \bibnamefont {Blaker}}, \ and\
  \bibinfo {author} {\bibfnamefont {A.~R.}\ \bibnamefont {Boccaccini}},\
  }\bibfield  {title} {\enquote {\bibinfo {title} {{Biodegradable and bioactive
  porous polymer / inorganic composite scaffolds for bone tissue
  engineering}},}\ }\href@noop {} {\bibfield  {journal} {\bibinfo  {journal}
  {Biomaterials}\ }\textbf {\bibinfo {volume} {27}},\ \bibinfo {pages} {3413}
  (\bibinfo {year} {2006})}\BibitemShut {NoStop}%
\bibitem [{\citenamefont {Hollister}(2005)}]{Hollister2005}%
  \BibitemOpen
  \bibfield  {author} {\bibinfo {author} {\bibfnamefont {S.~J.}\ \bibnamefont
  {Hollister}},\ }\bibfield  {title} {\enquote {\bibinfo {title} {{Porous
  scaffold design for tissue engineering}},}\ }\href@noop {} {\bibfield
  {journal} {\bibinfo  {journal} {Nat. Mater.}\ }\textbf {\bibinfo {volume}
  {4}},\ \bibinfo {pages} {518} (\bibinfo {year} {2005})}\BibitemShut {NoStop}%
\bibitem [{\citenamefont {Khan}\ \emph {et~al.}(2015)\citenamefont {Khan},
  \citenamefont {Tanaka},\ and\ \citenamefont {Ahmad}}]{Khan2015a}%
  \BibitemOpen
  \bibfield  {author} {\bibinfo {author} {\bibfnamefont {F.}~\bibnamefont
  {Khan}}, \bibinfo {author} {\bibfnamefont {M.}~\bibnamefont {Tanaka}}, \ and\
  \bibinfo {author} {\bibfnamefont {S.~R.}\ \bibnamefont {Ahmad}},\ }\bibfield
  {title} {\enquote {\bibinfo {title} {{Fabrication of polymeric biomaterials:
  a strategy for tissue engineering and medical devices}},}\ }\href@noop {}
  {\bibfield  {journal} {\bibinfo  {journal} {J. Mater. Chem. B}\ }\textbf
  {\bibinfo {volume} {3}},\ \bibinfo {pages} {8224} (\bibinfo {year}
  {2015})}\BibitemShut {NoStop}%
\bibitem [{\citenamefont {Roseti}\ \emph {et~al.}(2017)\citenamefont {Roseti},
  \citenamefont {Parisi}, \citenamefont {Petretta}, \citenamefont {Cavallo},
  \citenamefont {Desando}, \citenamefont {Bartolotti},\ and\ \citenamefont
  {Grigolo}}]{Roseti2017}%
  \BibitemOpen
  \bibfield  {author} {\bibinfo {author} {\bibfnamefont {L.}~\bibnamefont
  {Roseti}}, \bibinfo {author} {\bibfnamefont {V.}~\bibnamefont {Parisi}},
  \bibinfo {author} {\bibfnamefont {M.}~\bibnamefont {Petretta}}, \bibinfo
  {author} {\bibfnamefont {C.}~\bibnamefont {Cavallo}}, \bibinfo {author}
  {\bibfnamefont {G.}~\bibnamefont {Desando}}, \bibinfo {author} {\bibfnamefont
  {I.}~\bibnamefont {Bartolotti}}, \ and\ \bibinfo {author} {\bibfnamefont
  {B.}~\bibnamefont {Grigolo}},\ }\bibfield  {title} {\enquote {\bibinfo
  {title} {{Scaffolds for Bone Tissue Engineering : State of the art and new
  perspectives}},}\ }\href@noop {} {\bibfield  {journal} {\bibinfo  {journal}
  {Mater. Sci. Eng. C}\ }\textbf {\bibinfo {volume} {78}},\ \bibinfo {pages}
  {1246} (\bibinfo {year} {2017})}\BibitemShut {NoStop}%
\bibitem [{\citenamefont {O'Brien}(2011)}]{OBrien2011}%
  \BibitemOpen
  \bibfield  {author} {\bibinfo {author} {\bibfnamefont {F.~J.}\ \bibnamefont
  {O'Brien}},\ }\bibfield  {title} {\enquote {\bibinfo {title} {{Biomaterials
  {\&} scaffolds for tissue engineering}},}\ }\href@noop {} {\bibfield
  {journal} {\bibinfo  {journal} {Materials Today}\ }\textbf {\bibinfo {volume}
  {14}},\ \bibinfo {pages} {88} (\bibinfo {year} {2011})}\BibitemShut {NoStop}%
\bibitem [{\citenamefont {Neto}\ \emph {et~al.}(2019)\citenamefont {Neto},
  \citenamefont {Oliveira},\ and\ \citenamefont {Mano}}]{Neto2019}%
  \BibitemOpen
  \bibfield  {author} {\bibinfo {author} {\bibfnamefont {M.~D.}\ \bibnamefont
  {Neto}}, \bibinfo {author} {\bibfnamefont {M.~B.}\ \bibnamefont {Oliveira}},
  \ and\ \bibinfo {author} {\bibfnamefont {J.~F.}\ \bibnamefont {Mano}},\
  }\bibfield  {title} {\enquote {\bibinfo {title} {{Microparticles in Contact
  with Cells: From Carriers to Multifunctional Tissue Modulators}},}\ }\href
  {\doibase 10.1016/j.tibtech.2019.02.008} {\bibfield  {journal} {\bibinfo
  {journal} {Trends in Biotechnology}\ }\textbf {\bibinfo {volume} {37}},\
  \bibinfo {pages} {1011} (\bibinfo {year} {2019})}\BibitemShut {NoStop}%
\bibitem [{\citenamefont {Temenoff}\ and\ \citenamefont
  {Mikos}(2000)}]{Temenoff2000}%
  \BibitemOpen
  \bibfield  {author} {\bibinfo {author} {\bibfnamefont {J.~S.}\ \bibnamefont
  {Temenoff}}\ and\ \bibinfo {author} {\bibfnamefont {A.~G.}\ \bibnamefont
  {Mikos}},\ }\bibfield  {title} {\enquote {\bibinfo {title} {{Injectable
  biodegradable materials for orthopedic tissue engineering}},}\ }\href@noop {}
  {\bibfield  {journal} {\bibinfo  {journal} {Biomaterials}\ }\textbf {\bibinfo
  {volume} {21}},\ \bibinfo {pages} {2405} (\bibinfo {year}
  {2000})}\BibitemShut {NoStop}%
\bibitem [{\citenamefont {Oliveira}\ and\ \citenamefont
  {Mano}(2011)}]{Oliveira2011}%
  \BibitemOpen
  \bibfield  {author} {\bibinfo {author} {\bibfnamefont {M.~B.}\ \bibnamefont
  {Oliveira}}\ and\ \bibinfo {author} {\bibfnamefont {J.~F.}\ \bibnamefont
  {Mano}},\ }\bibfield  {title} {\enquote {\bibinfo {title} {{Polymer-based
  microparticles in tissue engineering and regenerative medicine}},}\
  }\href@noop {} {\bibfield  {journal} {\bibinfo  {journal} {Biotechnol.
  Prog.}\ }\textbf {\bibinfo {volume} {27}},\ \bibinfo {pages} {897} (\bibinfo
  {year} {2011})}\BibitemShut {NoStop}%
\bibitem [{\citenamefont {Cust{\'{o}}dio}\ \emph
  {et~al.}(2014{\natexlab{a}})\citenamefont {Cust{\'{o}}dio}, \citenamefont
  {Santo}, \citenamefont {Oliveira}, \citenamefont {Gomes}, \citenamefont
  {Reis},\ and\ \citenamefont {Mano}}]{Custodio2014}%
  \BibitemOpen
  \bibfield  {author} {\bibinfo {author} {\bibfnamefont {C.~A.}\ \bibnamefont
  {Cust{\'{o}}dio}}, \bibinfo {author} {\bibfnamefont {V.~E.}\ \bibnamefont
  {Santo}}, \bibinfo {author} {\bibfnamefont {M.~B.}\ \bibnamefont {Oliveira}},
  \bibinfo {author} {\bibfnamefont {M.~E.}\ \bibnamefont {Gomes}}, \bibinfo
  {author} {\bibfnamefont {R.~L.}\ \bibnamefont {Reis}}, \ and\ \bibinfo
  {author} {\bibfnamefont {J.~F.}\ \bibnamefont {Mano}},\ }\bibfield  {title}
  {\enquote {\bibinfo {title} {{Functionalized microparticles producing
  scaffolds in combination with cells}},}\ }\href@noop {} {\bibfield  {journal}
  {\bibinfo  {journal} {Adv. Funct. Mater.}\ }\textbf {\bibinfo {volume}
  {24}},\ \bibinfo {pages} {1391} (\bibinfo {year}
  {2014}{\natexlab{a}})}\BibitemShut {NoStop}%
\bibitem [{\citenamefont {Cust{\'{o}}dio}\ \emph {et~al.}(2015)\citenamefont
  {Cust{\'{o}}dio}, \citenamefont {Cerqueira}, \citenamefont {Marques},
  \citenamefont {Reis},\ and\ \citenamefont {Mano}}]{Custodio2015}%
  \BibitemOpen
  \bibfield  {author} {\bibinfo {author} {\bibfnamefont {C.~A.}\ \bibnamefont
  {Cust{\'{o}}dio}}, \bibinfo {author} {\bibfnamefont {M.~T.}\ \bibnamefont
  {Cerqueira}}, \bibinfo {author} {\bibfnamefont {A.~P.}\ \bibnamefont
  {Marques}}, \bibinfo {author} {\bibfnamefont {R.~L.}\ \bibnamefont {Reis}}, \
  and\ \bibinfo {author} {\bibfnamefont {J.~F.}\ \bibnamefont {Mano}},\
  }\bibfield  {title} {\enquote {\bibinfo {title} {{Cell selective chitosan
  microparticles as injectable cell carriers for tissue regeneration}},}\
  }\href@noop {} {\bibfield  {journal} {\bibinfo  {journal} {Biomaterials}\
  }\textbf {\bibinfo {volume} {43}},\ \bibinfo {pages} {23} (\bibinfo {year}
  {2015})}\BibitemShut {NoStop}%
\bibitem [{\citenamefont {Peng}\ \emph {et~al.}(2016)\citenamefont {Peng},
  \citenamefont {Kroes-Nijboer}, \citenamefont {Venema},\ and\ \citenamefont
  {van~der Linden}}]{Peng2016}%
  \BibitemOpen
  \bibfield  {author} {\bibinfo {author} {\bibfnamefont {J.}~\bibnamefont
  {Peng}}, \bibinfo {author} {\bibfnamefont {A.}~\bibnamefont {Kroes-Nijboer}},
  \bibinfo {author} {\bibfnamefont {P.}~\bibnamefont {Venema}}, \ and\ \bibinfo
  {author} {\bibfnamefont {E.}~\bibnamefont {van~der Linden}},\ }\bibfield
  {title} {\enquote {\bibinfo {title} {{Stability of colloidal dispersions in
  the presence of protein fibrils}},}\ }\href@noop {} {\bibfield  {journal}
  {\bibinfo  {journal} {Soft Matt.}\ }\textbf {\bibinfo {volume} {12}},\
  \bibinfo {pages} {3514} (\bibinfo {year} {2016})}\BibitemShut {NoStop}%
\bibitem [{\citenamefont {Joshi}\ \emph {et~al.}(2016)\citenamefont {Joshi},
  \citenamefont {Bargteil}, \citenamefont {Caciagli}, \citenamefont
  {Burelbach}, \citenamefont {Xing}, \citenamefont {Nunes}, \citenamefont
  {Pinto}, \citenamefont {Ara{\'{u}}jo}, \citenamefont {Bruijc},\ and\
  \citenamefont {Eiser}}]{Joshi2016}%
  \BibitemOpen
  \bibfield  {author} {\bibinfo {author} {\bibfnamefont {D.}~\bibnamefont
  {Joshi}}, \bibinfo {author} {\bibfnamefont {D.}~\bibnamefont {Bargteil}},
  \bibinfo {author} {\bibfnamefont {A.}~\bibnamefont {Caciagli}}, \bibinfo
  {author} {\bibfnamefont {J.}~\bibnamefont {Burelbach}}, \bibinfo {author}
  {\bibfnamefont {Z.}~\bibnamefont {Xing}}, \bibinfo {author} {\bibfnamefont
  {A.~S.}\ \bibnamefont {Nunes}}, \bibinfo {author} {\bibfnamefont {D.~E.~P.}\
  \bibnamefont {Pinto}}, \bibinfo {author} {\bibfnamefont {N.~A.~M.}\
  \bibnamefont {Ara{\'{u}}jo}}, \bibinfo {author} {\bibfnamefont
  {J.}~\bibnamefont {Bruijc}}, \ and\ \bibinfo {author} {\bibfnamefont
  {E.}~\bibnamefont {Eiser}},\ }\bibfield  {title} {\enquote {\bibinfo {title}
  {{Kinetic control of the coverage of oil droplets by DNA-functionalised
  colloids}},}\ }\href@noop {} {\bibfield  {journal} {\bibinfo  {journal} {Sci.
  Adv.}\ }\textbf {\bibinfo {volume} {2}},\ \bibinfo {pages} {e1600881}
  (\bibinfo {year} {2016})}\BibitemShut {NoStop}%
\bibitem [{\citenamefont {Antunes}\ \emph {et~al.}(2019)\citenamefont
  {Antunes}, \citenamefont {Dias}, \citenamefont {{Telo Da Gama}},\ and\
  \citenamefont {Ara{\'{u}}jo}}]{Antunes2019}%
  \BibitemOpen
  \bibfield  {author} {\bibinfo {author} {\bibfnamefont {G.~C.}\ \bibnamefont
  {Antunes}}, \bibinfo {author} {\bibfnamefont {C.~S.}\ \bibnamefont {Dias}},
  \bibinfo {author} {\bibfnamefont {M.~M.}\ \bibnamefont {{Telo Da Gama}}}, \
  and\ \bibinfo {author} {\bibfnamefont {N.~A.~M.}\ \bibnamefont
  {Ara{\'{u}}jo}},\ }\bibfield  {title} {\enquote {\bibinfo {title} {{Optimal
  number of linkers per monomer in linker-mediated aggregation}},}\ }\href@noop
  {} {\bibfield  {journal} {\bibinfo  {journal} {Soft Matt.}\ }\textbf
  {\bibinfo {volume} {15}},\ \bibinfo {pages} {3712} (\bibinfo {year}
  {2019})}\BibitemShut {NoStop}%
\bibitem [{\citenamefont {M{\"{u}}ller}\ \emph {et~al.}(2014)\citenamefont
  {M{\"{u}}ller}, \citenamefont {Bruinsma}, \citenamefont {Lieleg},
  \citenamefont {Bausch}, \citenamefont {Wall},\ and\ \citenamefont
  {Levine}}]{Muller2014}%
  \BibitemOpen
  \bibfield  {author} {\bibinfo {author} {\bibfnamefont {K.~W.}\ \bibnamefont
  {M{\"{u}}ller}}, \bibinfo {author} {\bibfnamefont {R.~F.}\ \bibnamefont
  {Bruinsma}}, \bibinfo {author} {\bibfnamefont {O.}~\bibnamefont {Lieleg}},
  \bibinfo {author} {\bibfnamefont {A.~R.}\ \bibnamefont {Bausch}}, \bibinfo
  {author} {\bibfnamefont {W.~A.}\ \bibnamefont {Wall}}, \ and\ \bibinfo
  {author} {\bibfnamefont {A.~J.}\ \bibnamefont {Levine}},\ }\bibfield  {title}
  {\enquote {\bibinfo {title} {{Rheology of semiflexible bundle networks with
  transient linkers}},}\ }\href@noop {} {\bibfield  {journal} {\bibinfo
  {journal} {Phys. Rev. Lett.}\ }\textbf {\bibinfo {volume} {112}},\ \bibinfo
  {pages} {238102} (\bibinfo {year} {2014})}\BibitemShut {NoStop}%
\bibitem [{\citenamefont {Bharti}\ \emph {et~al.}(2014)\citenamefont {Bharti},
  \citenamefont {Meissner}, \citenamefont {Klapp},\ and\ \citenamefont
  {Findenegg}}]{Bharti2014}%
  \BibitemOpen
  \bibfield  {author} {\bibinfo {author} {\bibfnamefont {B}~\bibnamefont
  {Bharti}}, \bibinfo {author} {\bibfnamefont {J}~\bibnamefont {Meissner}},
  \bibinfo {author} {\bibfnamefont {S~H~L}\ \bibnamefont {Klapp}}, \ and\
  \bibinfo {author} {\bibfnamefont {G~H}\ \bibnamefont {Findenegg}},\
  }\bibfield  {title} {\enquote {\bibinfo {title} {{Bridging interactions of
  proteins with silica nanoparticles: the influence of pH, ionic strength and
  protein concentration.}}}\ }\href@noop {} {\bibfield  {journal} {\bibinfo
  {journal} {Soft Matter}\ }\textbf {\bibinfo {volume} {10}},\ \bibinfo {pages}
  {718} (\bibinfo {year} {2014})}\BibitemShut {NoStop}%
\bibitem [{\citenamefont {Lindquist}\ \emph {et~al.}(2016)\citenamefont
  {Lindquist}, \citenamefont {Jadrich}, \citenamefont {Milliron},\ and\
  \citenamefont {Truskett}}]{Lindquist2016}%
  \BibitemOpen
  \bibfield  {author} {\bibinfo {author} {\bibfnamefont {B.~A.}\ \bibnamefont
  {Lindquist}}, \bibinfo {author} {\bibfnamefont {R.~B.}\ \bibnamefont
  {Jadrich}}, \bibinfo {author} {\bibfnamefont {D.~J.}\ \bibnamefont
  {Milliron}}, \ and\ \bibinfo {author} {\bibfnamefont {T.~M.}\ \bibnamefont
  {Truskett}},\ }\bibfield  {title} {\enquote {\bibinfo {title} {{On the
  formation of equilibrium gels via a macroscopic bond limitation}},}\
  }\href@noop {} {\bibfield  {journal} {\bibinfo  {journal} {J. Chem. Phys.}\
  }\textbf {\bibinfo {volume} {145}},\ \bibinfo {pages} {074906} (\bibinfo
  {year} {2016})}\BibitemShut {NoStop}%
\bibitem [{\citenamefont {Singh}\ \emph {et~al.}(2015)\citenamefont {Singh},
  \citenamefont {Lindquist}, \citenamefont {Ong}, \citenamefont {Jadrich},
  \citenamefont {Singh}, \citenamefont {Ha}, \citenamefont {Ellison},
  \citenamefont {Truskett},\ and\ \citenamefont {Milliron}}]{Singh2015}%
  \BibitemOpen
  \bibfield  {author} {\bibinfo {author} {\bibfnamefont {A.}~\bibnamefont
  {Singh}}, \bibinfo {author} {\bibfnamefont {B.~A.}\ \bibnamefont
  {Lindquist}}, \bibinfo {author} {\bibfnamefont {G.~K.}\ \bibnamefont {Ong}},
  \bibinfo {author} {\bibfnamefont {R.~B.}\ \bibnamefont {Jadrich}}, \bibinfo
  {author} {\bibfnamefont {A.}~\bibnamefont {Singh}}, \bibinfo {author}
  {\bibfnamefont {H.}~\bibnamefont {Ha}}, \bibinfo {author} {\bibfnamefont
  {C.~J.}\ \bibnamefont {Ellison}}, \bibinfo {author} {\bibfnamefont {T.~M.}\
  \bibnamefont {Truskett}}, \ and\ \bibinfo {author} {\bibfnamefont {D.~J.}\
  \bibnamefont {Milliron}},\ }\bibfield  {title} {\enquote {\bibinfo {title}
  {{Linking semiconductor nanocrystals into gel networks through all-inorganic
  bridges}},}\ }\href@noop {} {\bibfield  {journal} {\bibinfo  {journal}
  {Angew. Chem. Int. Ed.}\ }\textbf {\bibinfo {volume} {54}},\ \bibinfo {pages}
  {14840} (\bibinfo {year} {2015})}\BibitemShut {NoStop}%
\bibitem [{\citenamefont {Chen}\ \emph {et~al.}(2015)\citenamefont {Chen},
  \citenamefont {Kline},\ and\ \citenamefont {Liu}}]{Chen2015}%
  \BibitemOpen
  \bibfield  {author} {\bibinfo {author} {\bibfnamefont {J}~\bibnamefont
  {Chen}}, \bibinfo {author} {\bibfnamefont {S~R}\ \bibnamefont {Kline}}, \
  and\ \bibinfo {author} {\bibfnamefont {Y}~\bibnamefont {Liu}},\ }\bibfield
  {title} {\enquote {\bibinfo {title} {{From the depletion attraction to the
  bridging attraction: The effect of solvent molecules on the effective
  colloidal interactions}},}\ }\href@noop {} {\bibfield  {journal} {\bibinfo
  {journal} {J. Chem. Phys.}\ }\textbf {\bibinfo {volume} {142}},\ \bibinfo
  {pages} {84904} (\bibinfo {year} {2015})}\BibitemShut {NoStop}%
\bibitem [{\citenamefont {Luo}\ \emph {et~al.}(2015)\citenamefont {Luo},
  \citenamefont {Yuan}, \citenamefont {Zhao}, \citenamefont {Han},
  \citenamefont {Chen},\ and\ \citenamefont {Liu}}]{Luo2015}%
  \BibitemOpen
  \bibfield  {author} {\bibinfo {author} {\bibfnamefont {J.}~\bibnamefont
  {Luo}}, \bibinfo {author} {\bibfnamefont {G.}~\bibnamefont {Yuan}}, \bibinfo
  {author} {\bibfnamefont {C.}~\bibnamefont {Zhao}}, \bibinfo {author}
  {\bibfnamefont {C.~C.}\ \bibnamefont {Han}}, \bibinfo {author} {\bibfnamefont
  {J.}~\bibnamefont {Chen}}, \ and\ \bibinfo {author} {\bibfnamefont
  {Y.}~\bibnamefont {Liu}},\ }\bibfield  {title} {\enquote {\bibinfo {title}
  {{Gelation of large hard particles with short-range attraction induced by
  bridging of small soft microgels}},}\ }\href@noop {} {\bibfield  {journal}
  {\bibinfo  {journal} {Soft Matter}\ }\textbf {\bibinfo {volume} {11}},\
  \bibinfo {pages} {2494} (\bibinfo {year} {2015})}\BibitemShut {NoStop}%
\bibitem [{\citenamefont {Lowensohn}\ \emph {et~al.}(2019)\citenamefont
  {Lowensohn}, \citenamefont {Oyarz{\'{u}}n}, \citenamefont {{Narv{\'{a}}ez
  Paliza}}, \citenamefont {Mognetti},\ and\ \citenamefont
  {Rogers}}]{Lowensohn2019}%
  \BibitemOpen
  \bibfield  {author} {\bibinfo {author} {\bibfnamefont {J.}~\bibnamefont
  {Lowensohn}}, \bibinfo {author} {\bibfnamefont {B.}~\bibnamefont
  {Oyarz{\'{u}}n}}, \bibinfo {author} {\bibfnamefont {G.}~\bibnamefont
  {{Narv{\'{a}}ez Paliza}}}, \bibinfo {author} {\bibfnamefont {B.~M.}\
  \bibnamefont {Mognetti}}, \ and\ \bibinfo {author} {\bibfnamefont {W.~B.}\
  \bibnamefont {Rogers}},\ }\bibfield  {title} {\enquote {\bibinfo {title}
  {{Linker-Mediated Phase Behavior of DNA-Coated Colloids}},}\ }\href@noop {}
  {\bibfield  {journal} {\bibinfo  {journal} {Phys. Rev. X}\ }\textbf {\bibinfo
  {volume} {9}},\ \bibinfo {pages} {41054} (\bibinfo {year}
  {2019})}\BibitemShut {NoStop}%
\bibitem [{\citenamefont {Cyron}\ \emph {et~al.}(2013)\citenamefont {Cyron},
  \citenamefont {M{\"{u}}ller}, \citenamefont {Schmoller}, \citenamefont
  {Bausch}, \citenamefont {Wall},\ and\ \citenamefont {Bruinsma}}]{Cyron2013}%
  \BibitemOpen
  \bibfield  {author} {\bibinfo {author} {\bibfnamefont {C.~J.}\ \bibnamefont
  {Cyron}}, \bibinfo {author} {\bibfnamefont {K.~W.}\ \bibnamefont
  {M{\"{u}}ller}}, \bibinfo {author} {\bibfnamefont {K.~M.}\ \bibnamefont
  {Schmoller}}, \bibinfo {author} {\bibfnamefont {A.~R.}\ \bibnamefont
  {Bausch}}, \bibinfo {author} {\bibfnamefont {W.~A.}\ \bibnamefont {Wall}}, \
  and\ \bibinfo {author} {\bibfnamefont {R.~F.}\ \bibnamefont {Bruinsma}},\
  }\bibfield  {title} {\enquote {\bibinfo {title} {{Equilibrium phase diagram
  of semi-flexible polymer networks with linkers}},}\ }\href@noop {} {\bibfield
   {journal} {\bibinfo  {journal} {EPL}\ }\textbf {\bibinfo {volume} {102}},\
  \bibinfo {pages} {38003} (\bibinfo {year} {2013})}\BibitemShut {NoStop}%
\bibitem [{\citenamefont {Winer}\ \emph {et~al.}(2011)\citenamefont {Winer},
  \citenamefont {Chopra}, \citenamefont {Kresh},\ and\ \citenamefont
  {Janmey}}]{Johnson2011}%
  \BibitemOpen
  \bibfield  {author} {\bibinfo {author} {\bibfnamefont {J.~P.}\ \bibnamefont
  {Winer}}, \bibinfo {author} {\bibfnamefont {A.}~\bibnamefont {Chopra}},
  \bibinfo {author} {\bibfnamefont {J.~Y.}\ \bibnamefont {Kresh}}, \ and\
  \bibinfo {author} {\bibfnamefont {P.~A.}\ \bibnamefont {Janmey}},\ }\bibfield
   {title} {\enquote {\bibinfo {title} {{Substrate Elasticity as a Probe to
  Measure Mechanosensing at Cell-Cell and Cell-Matrix Junctions}},}\ }in\ \href
  {\doibase 10.1007/978-1-4419-8083-0} {\emph {\bibinfo {booktitle}
  {Mechanobiology of Cell-Cell and Cell-Matrix Interactions}}},\ \bibinfo
  {editor} {edited by\ \bibinfo {editor} {\bibfnamefont {A.~W.}\ \bibnamefont
  {Johnson}}\ and\ \bibinfo {editor} {\bibfnamefont {B.~A.~C.}\ \bibnamefont
  {Harley}}}\ (\bibinfo  {publisher} {Springer Science+Business Media, LLC},\
  \bibinfo {year} {2011})\ Chap.~\bibinfo {chapter} {2}, p.~\bibinfo {pages}
  {11}\BibitemShut {NoStop}%
\bibitem [{\citenamefont {Badami}\ \emph {et~al.}(2006)\citenamefont {Badami},
  \citenamefont {Kreke}, \citenamefont {Thompson}, \citenamefont {Riffle},\
  and\ \citenamefont {Goldstein}}]{Badami2006}%
  \BibitemOpen
  \bibfield  {author} {\bibinfo {author} {\bibfnamefont {A.~S.}\ \bibnamefont
  {Badami}}, \bibinfo {author} {\bibfnamefont {M.~R.}\ \bibnamefont {Kreke}},
  \bibinfo {author} {\bibfnamefont {M.~S.}\ \bibnamefont {Thompson}}, \bibinfo
  {author} {\bibfnamefont {J.~S.}\ \bibnamefont {Riffle}}, \ and\ \bibinfo
  {author} {\bibfnamefont {A.~S.}\ \bibnamefont {Goldstein}},\ }\bibfield
  {title} {\enquote {\bibinfo {title} {{Effect of fiber diameter on spreading ,
  proliferation , and differentiation of osteoblastic cells on electrospun poly
  ( lactic acid ) substrates}},}\ }\href@noop {} {\bibfield  {journal}
  {\bibinfo  {journal} {Biomaterials}\ }\textbf {\bibinfo {volume} {27}},\
  \bibinfo {pages} {596} (\bibinfo {year} {2006})}\BibitemShut {NoStop}%
\bibitem [{\citenamefont {Chen}\ \emph {et~al.}(2007)\citenamefont {Chen},
  \citenamefont {Patra}, \citenamefont {Warner},\ and\ \citenamefont
  {Bhowmick}}]{Chen2007}%
  \BibitemOpen
  \bibfield  {author} {\bibinfo {author} {\bibfnamefont {M.}~\bibnamefont
  {Chen}}, \bibinfo {author} {\bibfnamefont {P.~K.}\ \bibnamefont {Patra}},
  \bibinfo {author} {\bibfnamefont {S.~B.}\ \bibnamefont {Warner}}, \ and\
  \bibinfo {author} {\bibfnamefont {S.}~\bibnamefont {Bhowmick}},\ }\bibfield
  {title} {\enquote {\bibinfo {title} {{Role of Fiber Diameter in Adhesion and
  Proliferation of NIH 3T3 Fibroblast on Electrospun Polycaprolactone
  Scaffolds}},}\ }\href@noop {} {\bibfield  {journal} {\bibinfo  {journal}
  {Tissue Eng.}\ }\textbf {\bibinfo {volume} {13}},\ \bibinfo {pages} {579}
  (\bibinfo {year} {2007})}\BibitemShut {NoStop}%
\bibitem [{\citenamefont {Li}\ \emph {et~al.}(2001)\citenamefont {Li},
  \citenamefont {Laurencin}, \citenamefont {Caterson}, \citenamefont {Tuan},\
  and\ \citenamefont {Ko}}]{Li2001}%
  \BibitemOpen
  \bibfield  {author} {\bibinfo {author} {\bibfnamefont {W.-J.}\ \bibnamefont
  {Li}}, \bibinfo {author} {\bibfnamefont {C.~T.}\ \bibnamefont {Laurencin}},
  \bibinfo {author} {\bibfnamefont {E.~J.}\ \bibnamefont {Caterson}}, \bibinfo
  {author} {\bibfnamefont {R.~S.}\ \bibnamefont {Tuan}}, \ and\ \bibinfo
  {author} {\bibfnamefont {F.~K.}\ \bibnamefont {Ko}},\ }\bibfield  {title}
  {\enquote {\bibinfo {title} {{Electrospun nanofibrous structure : A novel
  scaffold for tissue engineering}},}\ }\href@noop {} {\bibfield  {journal}
  {\bibinfo  {journal} {J. Biomed. Mater. Res.}\ }\textbf {\bibinfo {volume}
  {60}},\ \bibinfo {pages} {613} (\bibinfo {year} {2001})}\BibitemShut
  {NoStop}%
\bibitem [{\citenamefont {Oliveira}\ and\ \citenamefont
  {Mano}(2014)}]{Oliveira2014a}%
  \BibitemOpen
  \bibfield  {author} {\bibinfo {author} {\bibfnamefont {M.~B.}\ \bibnamefont
  {Oliveira}}\ and\ \bibinfo {author} {\bibfnamefont {J.~F.}\ \bibnamefont
  {Mano}},\ }\bibfield  {title} {\enquote {\bibinfo {title} {{High-throughput
  screening for integrative biomaterials design: Exploring advances and new
  trends}},}\ }\href {\doibase 10.1016/j.tibtech.2014.09.009} {\bibfield
  {journal} {\bibinfo  {journal} {Trends in Biotechnology}\ }\textbf {\bibinfo
  {volume} {32}},\ \bibinfo {pages} {627} (\bibinfo {year} {2014})}\BibitemShut
  {NoStop}%
\bibitem [{\citenamefont {Cust{\'{o}}dio}\ \emph
  {et~al.}(2014{\natexlab{b}})\citenamefont {Cust{\'{o}}dio}, \citenamefont
  {Reis},\ and\ \citenamefont {Mano}}]{Custodio2014a}%
  \BibitemOpen
  \bibfield  {author} {\bibinfo {author} {\bibfnamefont {C.~A.}\ \bibnamefont
  {Cust{\'{o}}dio}}, \bibinfo {author} {\bibfnamefont {R.~L.}\ \bibnamefont
  {Reis}}, \ and\ \bibinfo {author} {\bibfnamefont {J.~F.}\ \bibnamefont
  {Mano}},\ }\bibfield  {title} {\enquote {\bibinfo {title} {{Engineering
  Biomolecular Microenvironments for Cell Instructive Biomaterials}},}\ }\href
  {\doibase 10.1002/adhm.201300603} {\bibfield  {journal} {\bibinfo  {journal}
  {Adv. Healthcare Mater.}\ }\textbf {\bibinfo {volume} {3}},\ \bibinfo {pages}
  {797} (\bibinfo {year} {2014}{\natexlab{b}})}\BibitemShut {NoStop}%
\bibitem [{\citenamefont {van~der Walt}\ \emph {et~al.}(2014)\citenamefont
  {van~der Walt}, \citenamefont {Sch{\"{o}}nberger}, \citenamefont
  {Nunez-Iglesias}, \citenamefont {Boulogne}, \citenamefont {Warner},
  \citenamefont {Yager}, \citenamefont {Gouillart}, \citenamefont {Yu},\ and\
  \citenamefont {the scikit-image Contributors}}]{VanderWalt2014}%
  \BibitemOpen
  \bibfield  {author} {\bibinfo {author} {\bibfnamefont {S.}~\bibnamefont
  {van~der Walt}}, \bibinfo {author} {\bibfnamefont {J.~L.}\ \bibnamefont
  {Sch{\"{o}}nberger}}, \bibinfo {author} {\bibfnamefont {J.}~\bibnamefont
  {Nunez-Iglesias}}, \bibinfo {author} {\bibfnamefont {F.}~\bibnamefont
  {Boulogne}}, \bibinfo {author} {\bibfnamefont {J.~D.}\ \bibnamefont
  {Warner}}, \bibinfo {author} {\bibfnamefont {N.}~\bibnamefont {Yager}},
  \bibinfo {author} {\bibfnamefont {E.}~\bibnamefont {Gouillart}}, \bibinfo
  {author} {\bibfnamefont {T}~\bibnamefont {Yu}}, \ and\ \bibinfo {author}
  {\bibnamefont {the scikit-image Contributors}},\ }\bibfield  {title}
  {\enquote {\bibinfo {title} {{scikit-image : image processing in Python}},}\
  }\href@noop {} {\bibfield  {journal} {\bibinfo  {journal} {PeerJ}\ }\textbf
  {\bibinfo {volume} {2}},\ \bibinfo {pages} {e453} (\bibinfo {year}
  {2014})}\BibitemShut {NoStop}%
\bibitem [{\citenamefont {Otsu}(1979)}]{Otsu1979}%
  \BibitemOpen
  \bibfield  {author} {\bibinfo {author} {\bibfnamefont {N}~\bibnamefont
  {Otsu}},\ }\bibfield  {title} {\enquote {\bibinfo {title} {{A Threshold
  Selection Method from Gray-Level Histograms}},}\ }\href@noop {} {\bibfield
  {journal} {\bibinfo  {journal} {IEEE Trans. Sys. Man. Cyber.}\ }\textbf
  {\bibinfo {volume} {9}},\ \bibinfo {pages} {62} (\bibinfo {year}
  {1979})}\BibitemShut {NoStop}%
\bibitem [{\citenamefont {Dias}\ \emph {et~al.}(2016)\citenamefont {Dias},
  \citenamefont {Braga}, \citenamefont {Ara{\'{u}}jo},\ and\ \citenamefont
  {{Telo da Gama}}}]{Dias2016}%
  \BibitemOpen
  \bibfield  {author} {\bibinfo {author} {\bibfnamefont {C.~S.}\ \bibnamefont
  {Dias}}, \bibinfo {author} {\bibfnamefont {C.}~\bibnamefont {Braga}},
  \bibinfo {author} {\bibfnamefont {N.~A.~M.}\ \bibnamefont {Ara{\'{u}}jo}}, \
  and\ \bibinfo {author} {\bibfnamefont {M.~M.}\ \bibnamefont {{Telo da
  Gama}}},\ }\bibfield  {title} {\enquote {\bibinfo {title} {{Relaxation
  dynamics of functionalized colloids on attractive substrates}},}\ }\href@noop
  {} {\bibfield  {journal} {\bibinfo  {journal} {Soft Matt.}\ }\textbf
  {\bibinfo {volume} {12}},\ \bibinfo {pages} {1550} (\bibinfo {year}
  {2016})}\BibitemShut {NoStop}%
\bibitem [{\citenamefont {Ara{\'{u}}jo}\ \emph {et~al.}(2017)\citenamefont
  {Ara{\'{u}}jo}, \citenamefont {Dias},\ and\ \citenamefont {{Telo da
  Gama}}}]{Araujo2017}%
  \BibitemOpen
  \bibfield  {author} {\bibinfo {author} {\bibfnamefont {N.~A.~M.}\
  \bibnamefont {Ara{\'{u}}jo}}, \bibinfo {author} {\bibfnamefont {C.~S.}\
  \bibnamefont {Dias}}, \ and\ \bibinfo {author} {\bibfnamefont {M.~M.}\
  \bibnamefont {{Telo da Gama}}},\ }\bibfield  {title} {\enquote {\bibinfo
  {title} {{Nonequilibrium self-organization of colloidal particles on
  substrates: adsorption, relaxation, and annealing}},}\ }\href@noop {}
  {\bibfield  {journal} {\bibinfo  {journal} {J. Phys.: Condens. Matter}\
  }\textbf {\bibinfo {volume} {29}},\ \bibinfo {pages} {014001} (\bibinfo
  {year} {2017})}\BibitemShut {NoStop}%
\bibitem [{\citenamefont {Plimpton}(1995)}]{Plimpton1995}%
  \BibitemOpen
  \bibfield  {author} {\bibinfo {author} {\bibfnamefont {S.}~\bibnamefont
  {Plimpton}},\ }\bibfield  {title} {\enquote {\bibinfo {title} {{Fast parallel
  algorithms for short-range Molecular Dynamics}},}\ }\href@noop {} {\bibfield
  {journal} {\bibinfo  {journal} {J. Comp. Phys.}\ }\textbf {\bibinfo {volume}
  {117}},\ \bibinfo {pages} {1} (\bibinfo {year} {1995})}\BibitemShut {NoStop}%
\end{thebibliography}%

\section{acknowledgments} We acknowledge financial support from the Portuguese
Foundation for Science and Technology (FCT) under Contracts no.
PTDC/FIS-MAC/28146/2017 (LISBOA-01-0145-FEDER-028146), UIDB/00618/2020,
UIDP/00618/2020, and CEECIND/00586/2017.
This work was also developed within the scope of the project CICECO-Aveiro Institute of Materials, UIDB/50011/2020 and UIDP/50011/2020, financed by national funds through the Portuguese Foundation for Science and Technology/MCTES. The authors also acknowledge the funding from the European Research Council (ERC) for project ATLAS (ERC-2014-ADG-669858).
 
\end{document}